# Quantum Threat in Healthcare IoT: Challenges and Mitigation Strategies

Asif Alif[1], Khondokar Fida Hasan[2*], Jesse Laeuchli[2], Mohammad Jabed Morshed Chowdhury[3]

## Abstract

The Internet of Things (IoT) has transformed healthcare, facilitating remote patient monitoring, enhanced medication adherence, and chronic disease management. However, this interconnected ecosystem faces significant vulnerabilities with the advent of quantum computing, which threatens to break existing encryption standards protecting sensitive patient data in IoT-enabled medical devices. This chapter examines the quantum threat to healthcare IoT security, highlighting the potential impacts of compromised encryption, including privacy breaches, device failures, and manipulated medical records. It introduces post-quantum cryptography (PQC) and quantum-resistant techniques like quantum key distribution (QKD), addressing their application in resource-constrained healthcare IoT devices such as pacemakers, monitoring tools, and telemedicine systems. The chapter further explores the challenges of integrating these solutions and reviews global efforts in mitigating quantum risks, offering insights into suitable PQC primitives for various healthcare use cases.

## Contents

1 Islamic University, Kushtia, Bangladesh; 2 University of New South Wales (UNSW), Canberra, Australia; 3 La Trobe University, Melbourne, Australia | *Corresponding Author Email: fida.hasan@unsw.edu.au

Submitted as a Book Chapter

# 1   Introduction

The healthcare sector has been significantly influenced by the digital revolution, leading to the emergence of the Digital Health era. One of the primary catalysts behind this paradigm shift is the Internet of Things (IoT), which facilitates the interconnection of medical equipment, sensors, and wearables to gather patient data in real-time. The interconnectivity facilitates the ability to remotely monitor patients, manage chronic diseases, and enhance drug adherence [1]. Nevertheless, the advantages of the Internet of Things (IoT) give rise to security concerns, which could undermine the security of sensitive health data [2].

The emergence of quantum computing poses a significant and imminent threat. Quantum machines utilise the principles of quantum mechanics to achieve significantly faster calculations compared to conventional computers [3]. The potential exists for the breach of existing encryption standards, which serve as the fundamental basis for ensuring data security in the healthcare Internet of Things (IoT) [4]. Although the general implementation of large-scale quantum computers may be delayed by a few years, it is essential to take early steps to guarantee the enduring security of critical patient data.

The cryptographic algorithms employed in healthcare IoT systems offer the CIA triad, which encompasses Confidentiality (ensuring the privacy of data), Integrity (ensuring that data remains unaltered), and Availability (ensuring authorised access to data) [3]. Nevertheless, the advent of quantum computing poses a potential threat to these protective measures. Quantum computers have the capacity to decrypt existing encrypted data, which could result in the exposure of patient health information, disruption of medical device operations, and the ability to manipulate healthcare records.

This chapter focuses on the crucial matter of ensuring the security of healthcare Internet of Things (IoT) devices in the quantum age. This study examines the risks presented by quantum computing and investigates the potential solutions of post-quantum cryptography (PQC) and other quantum-resistant approaches. This study examines the limits associated with conventional healthcare Internet of Things (IoT) devices, such as pacemakers, monitoring tools, and telemedicine equipment. Additionally, it investigates the potential adaptations of PQC primitives to address the unique resource limitations of these devices. We will go deeper into the current mitigation initiatives and present a strategic plan for ensuring the security of healthcare IoT in the era of quantum technology.

The chapter is organised in the following manner: In Section 2, an overview of the quantum threat to cryptography is presented. In section 3, a comprehensive examination of the quantum threat and PQC approaches is presented. Section 4 examines the cryptographic requirements of the healthcare Internet of Things (IoT) and investigates appropriate PQC primitives for various scenarios. Section 5 delves into the difficulties associated with implementing these strategies and investigates current mitigation initiatives on a global scale. Section 6 is concluded prior to the presentation of the reference.

## 2 An Overview of Quantum Threat to Cryptography:

This section briefly introduces quantum computing and its threat to popular cryptographic systems that are the cornerstone of today's digital world.

### 2.1 A Brief Introduction to Quantum Computers:

Quantum computing represents a revolutionary approach to processing information, leveraging the principle of quantum mechanics to solve problems that are intractable for classical computers. It employs the laws of quantum mechanics to perform complex computations in seconds that are too complex for classical computers. At the heart of this technology are qubits (as shown in the Figure 1), which, unlike the binary bits of classical computing that are either 'on' (1) or 'off' (0), can exist in both states simultaneously through a property called superposition.

In addition, qubits exhibit entanglement, a unique quantum condition where the state of one qubit is intrinsically linked to another, irrespective of distance. The combined capabilities of superposition and entanglement grant quantum computers their extraordinary potential, far surpassing the limits of classical computational power [5].

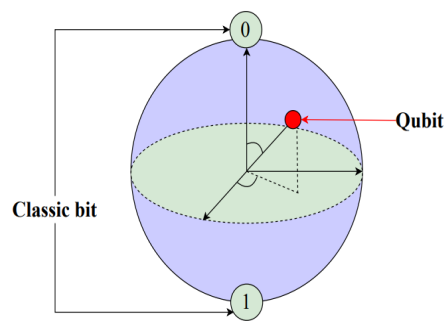

Figure-1: A simplified presentation of Qbit

### 2.2 Quantum Threat:

Quantum computing heralds a new frontier in technological advancement, promising unprecedented progress in fields ranging from medicine and chemistry to automated systems and artificial intelligence. Yet, this cutting-edge innovation brings with it a formidable challenge to cybersecurity: the quantum threat. Traditional encryption methods, which safeguard our most sensitive data, rely on the computational difficulty of tasks like factoring large numbers or calculating discrete logarithms—tasks that quantum computers could solve with alarming speed and efficiency.

Encryption, the cornerstone of secure communication, employs complex codes—keys and algorithms—to ensure that only authorised recipients can access and interpret the information. This protective measure is vital for maintaining the confidentiality and integrity of digital communications, including emails, passwords, and financial transactions.  The advent of quantum computing, however, could render these encryption systems obsolete, potentially laying bare the very secrets they were designed to protect.  This could be used by hackers or state actors to cause great disruption or for illegal means, such as stealing personal information, spying on communications, or sabotaging critical infrastructure. This is what we call the quantum threat, and it is one of the most urgent and important challenges for cryptography nowadays [6].

Cryptography, the science of secure communication, is built primarily on two foundational algorithm types. The Symmetric Key Algorithm, or single-key encryption, requires both sender and receiver to use identical keys for encrypting and decrypting messages. In contrast, the Asymmetric Key Algorithm employs a public key for encryption and a private key for decryption, with RSA being a prominent example in contemporary digital security [7].

Addressing the quantum threat is thus one of the most pressing and complex challenges facing cryptography today. It necessitates re-evaluating our current security paradigms and developing quantum-resistant cryptographic techniques to protect against the vulnerabilities exposed by quantum computing's rise.

### 2.3   Quantum Threats to Asymmetric Cryptography:

Asymmetric cryptography, also known as public-key cryptography, is a type of encryption that uses two different keys: a public key and a private key. The public key is used to encrypt messages and can be shared with anyone, while the private key is used to decrypt messages and must be kept secret. A widely used asymmetric cryptography algorithm in today's digital world is RSA, which stands for Rivest-Shamir-Adleman, the names of its inventors. RSA relies on the difficulty of factoring large semi-primes, which are numbers that are the product of two prime numbers, such as 15 = 3 x 5 [8]. The security of RSA depends on the size of the key, which is measured in bits. For example, RSA 1024 uses a 1024-bit key, which is equivalent to a 309-digit number. The larger the key, the harder it is to factor in and compromise the encryption [9, 10].

The security of the RSA algorithm solely depends on the computational cost of factoring two large numbers, a quantum computer with appropriate algorithms can compromise RSA encryption with ease.  For example, a quantum computer can use Shor's algorithm, discovered by Peter Shor in 1994, to factor large numbers much faster than classical computers. This poses a threat to RSA and other public-key encryption schemes. Shor's algorithm operates in polynomial time, meaning its computational time grows slowly with input size. In contrast, the best classical factoring algorithms work in exponential time. For instance, Shor's algorithm can factor a 1024-bit number in about 10 hours, while classical methods would take around 10 billion years [11].

To break RSA 1024, a quantum computer would need approximately 2,050 logical qubits. Logical qubits, which are error-corrected, enable reliable computations by encoding multiple physical qubits to protect against errors [12].

Currently, quantum computers are not yet capable of achieving this level of performance. However, advancements are ongoing. For example, IBM's 1,121-qubit processor, IBM Condor, introduced in December 2023, demonstrates significant progress towards achieving quantum supremacy [13-15].

Table 1 presents different kinds of today's cryptosystems with their category, key size, and security parameters. Quantum algorithm, logical Qubits, physical qubits, and time required to break specific key size of each cryptosystem are also presented in Table 1 [15].

Table-1: Quantum Resources and Estimated Time Required to Break Current Common Cryptosystems [15]

| Cryptosystem | Category | Key Size | Security Parameter | Quantum Algorithm | Requirement | | |
|---|---|---|---|---|---|---|---|
| | | | | | Logical Qubits | Physical Qubits | Time |
| RSA | Asymmetric Encryption | 1024 | 80 | Shor's algorithm | 2,050 | 8.05 × 10^6 | 3.58 hours |
| | | 2048 | 112 | | 4,098 | 8.56 × 10^6 | 28.63 hours |
| | | 4096 | 128 | | 8,194 | 1.12 × 10^7 | 229 hours |
| ECC Discrete log problem | Asymmetric Encryption | 256 | 128 | Shor's algorithm | 2,330 | 8.56 × 10^6 | 10.5 hours |
| | | 384 | 192 | | 3,484 | 9.05 × 10^6 | 37.67 hours |
| | | 521 | 256 | | 4,719 | 1.13 × 10^6 | 55 hours |
| AES-GCM | Symmetric Encryption | 128 | 128 | Grover's algorithm | 2,953 | 4.61 × 10^6 | 2.61 × 10^12 yrs |
| | | 192 | 192 | | 4,449 | 1.68 × 10^7 | 1.97× 10^22 yrs |
| | | 256 | 256 | | 6,681 | 3.36 × 10^7 | 2.29 × 10^32 yrs |
| SHA256 | Bitcoin mining | N/A | 72 | Grover's algorithm | 2,403 | 2.23 × 10^6 | 1.8 × 10^4 yrs |
| PBKDF2 with 1000 iterations | Password hashing | N/A | 66 | Grover's algorithm | 2,403 | 2.23 × 10^6 | 2.3 × 10^7 yrs |

More generally, quantum computers are capable of attacking any cryptosystem that is based on the hidden abelian subgroup problem [16, 17]. While RSA is one popular cryptosystem that reduces to this problem, other popular methods, such as those based on the elliptic curve discrete logarithm problem are also vulnerable.

## 2.4 Quantum Threat to Symmetric Cryptography:

Symmetric cryptography, or secret-key cryptography, uses the same key for both encryption and decryption. The key must be securely shared between the sender and receiver and kept secret from others. AES (Advanced Encryption Standard) is a popular symmetric algorithm, with its security depending on key size. For example, AES 128 uses a 128-bit key. The larger the key, the harder it is to find and break the encryption [18].

While symmetric cryptography is more resistant to quantum threats than asymmetric cryptography, quantum computers can still pose a risk. They offer a quadratic speedup in brute force attacks, reducing the effective security of keys. For instance, a quantum computer would need roughly 2^64 attempts to break a 128-bit key, halving its security. To maintain security against quantum attacks, larger keys, such as AES 256, are recommended [19].

This threat to symmetric key cryptography, arises from Grover's algorithm. This algorithm, discovered by Lov Grover in 1996, is a quantum algorithm that can search unsorted databases much faster than classical computers. It can find an item in a database of N items in about √N steps, compared to the classical average of N/2 steps. For example, Grover's algorithm can search a billion-item database in about 31,623 steps, while a classical algorithm would need about 500 million steps [20]. By treating a symmetric key encryption function, as such an unsorted search problem the attacker can obtain a similar quadratic speedup.

The quantum threat to cryptography is not only a future concern but also a current one. Quantum computers could retroactively break encryption by storing encrypted data today and decrypting it later when sufficiently powerful quantum computers are available. This means that data protected by current encryption methods could be vulnerable to future quantum attacks [20].

To estimate the time required to break symmetric cryptographic algorithms using Grover's algorithm, we'll need to consider the number of operations Grover's algorithm performs and the speed of quantum computers.

For simplicity, let's assume a hypothetical quantum computer can perform $10^9$ (1 billion) quantum operations per second then the time required to break a symmetric cryptographic algorithm is: Time (seconds) = Grover's Algorithm Complexity / Operations per Second and Time (years) = Time (seconds) / 60×60×24×365.

Table 2 provides a comprehensive analysis of the time required to break various cryptosystems using Grover's algorithm under the assumption of performing $10^9$ operations per second. The table includes symmetric encryption algorithms such as AES, DES, 3DES, Blowfish, and RC4, detailing their key sizes and the corresponding computational complexity with Grover's algorithm. This research emphasises the critical necessity for the development and implementation of quantum-resistant encryption solutions to strengthen cybersecurity in the age of quantum computing achievements.

Table-2: Quantum Attack Time Estimates for Various Symmetric Cryptosystems

| Cryptosystem | Category | Key Size | Quantum Algorithm | Complexity with Grover's Algorithm (operations) | Operations per second | Time (Approximately) |
|---|---|---|---|---|---|---|
| AES | Symmetric Encryption | 128 | Grover's algorithm | $O(2^{64})$ | $10^9$ | 585 yrs |
|  |  | 192 |  | $O(2^{96})$ |  | $2.51 \times 10^{11}$ yrs |
|  |  | 256 |  | $O(2^{128})$ |  | $1.08 \times 10^{21}$ yrs |
| DES | Symmetric Encryption | 56 | Grover's algorithm | $O(2^{28})$ | $10^9$ | Instant |
| 3DES | Symmetric Encryption | 112 | Grover's algorithm | $O(2^{56})$ | $10^9$ | 2.28 yrs |
|  |  | 168 |  | $O(2^{84})$ |  | $6.12 \times 10^8$ yrs |

| Blowfish | Symmetric Encryption | 32 | Grover's algorithm | O(2^16) | 10^9 | Instant |
| | | 448 | | O(2^224) | | 1.18 × 10^51 yrs |
| RC4 | Symmetric Encryption | 40 | Grover's algorithm | O(2^20) | 10^9 | Instant |
| | | 2048 | | O(2^1024) | | 1.69 × 10^300 yrs |

## 2.5 Quantum Threat to Hash Cryptography:

Hash cryptography serves as a cornerstone in contemporary security frameworks, facilitating crucial functionalities such as data integrity verification, digital signatures, and password hashing. Its operation revolves around converting input data, regardless of its size, into fixed-length hash values. This process guarantees that even minor alterations in the input yield markedly distinct hash outputs. Hash functions stand as pivotal constructs in cryptography, facilitating validation while preserving confidentiality. Consequently, they play a pivotal role in bolstering mechanisms for data authentication and integrity, exemplified by hash-based message authentication codes (HMAC) and digital signatures. Cryptographic hash functions (CHFs) provide efficient and secure methods for authentication and integrity verification without compromising confidentiality. They are crucial for tasks like software verification and user authentication in web applications. A useful CHF should satisfy several key properties like uniformity, determinism, irreversibility, approximate injectivity to validate a piece of data against the original instance by comparing a digest of the data to a digest of the original [21].

Cryptographic hash functions (CHFs) are evaluated according to how well they withstand collision and pre-image attacks as pre-image resistance makes it difficult to locate the original input given a hash, whereas collision resistance makes it difficult to locate two inputs that produce the same hash [21]. Both are essential for applications such as digital signatures to maintain security.

Brute force attacks are the principal quantum threat to cryptographic hashing. In order to find an input that yields the same digest, an attacker would experiment with different ones until they find one that works. $2^n$ possible values may be obtained using n bits in the input. To have a greater than 50% probability of success, the attacker must thus try out around $2^{(n-1)}$ inputs. Grover's approach applies quantum amplitude amplification to provide a quadratic speedup for such an unstructured search environment as this method increases the likelihood (amplitude) of the right answer when searching through a wide collection of options/definitions by utilising certain phenomena at the quantum interface. When this happens, a pre-image attack's time-based complexity drops to $2^{n/2}$ [21]. In practical terms, this means that a 256-bit CHF, which is currently considered secure against pre-image attacks by classical computers, would provide the same level of security as a 128-bit CHF when faced with a quantum attacker utilising Grover's algorithm.

BHT algorithm, a quantum algorithm proposed by Bassard, Høyer, and Tapp (BHT) in 1997 combines elements of the birthday attack with Grover search, offering a theoretical scaling of $O(2^{n/3})$ for finding hash collisions [22]. However, this improved scaling relies on the existence of quantum random access memory (QRAM) technology, which is not currently available. With ongoing research and advancements in quantum computing, there's a possibility that in the future, sufficient quantum random access memory (QRAM) may become available to effectively run the BHT algorithm. If realised, this scenario could pose a significant threat to Hash cryptographic security.

## 3   Introduction to PQC and other Quantum Resistant Techniques:

In anticipation of the quantum computing era, the development of quantum-resistant cryptographic methods has become eminent. These methods, also known as quantum-secure, post-quantum, or quantum-safe cryptography, are designed to thwart the advanced capabilities of quantum computers that could otherwise compromise traditional encryption techniques. Among these, post-quantum cryptography (PQC) stands out as a pivotal innovation. PQC encompasses algorithms that, while operational on conventional computers, are engineered to protect data against the formidable computational force of quantum attacks [23]. Such methods are based on problems that are known to be hard for quantum computers to solve. For example, while quantum computers can attack the abelian subgroup problem, they cannot solve such problems when the group is non-abelian, so crypto systems based on this problem should be secure. The National Institute of Standards and Technology (NIST) has been proactive in this domain, initiating a call for post-quantum algorithm proposals in 2016 to identify solutions resilient to quantum computing's prowess, with the standardisation process currently in its fourth round.

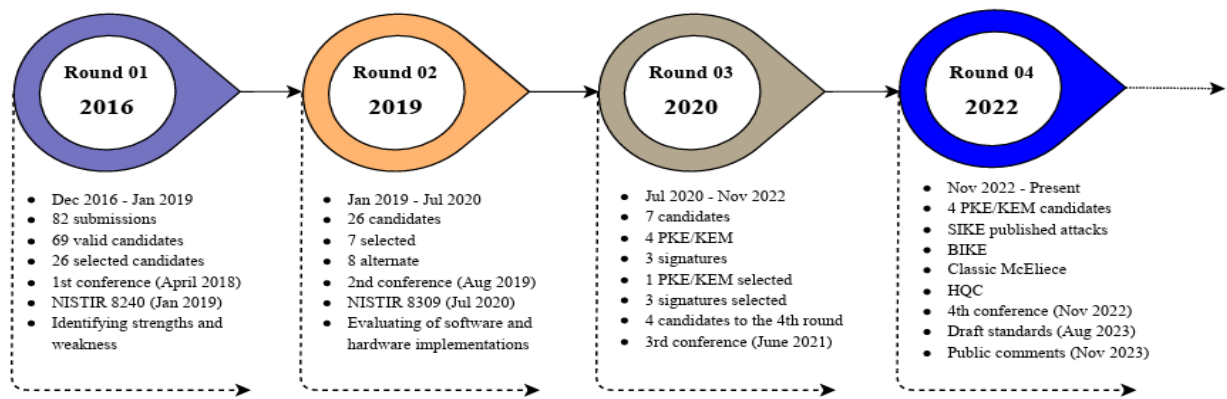

Figure-2: PQC Standardisation Process of NIST [24]

Figure 2 provides a concise visual representation of the NIST Post-Quantum Cryptography (PQC) standardisation process, delineating the four pivotal rounds undertaken by the National Institute of Standards and Technology (NIST). This figure serves as an illustrative guide, offering a comprehensive overview of the progressive stages through which cryptographic algorithms are rigorously evaluated and ultimately standardised to ensure resilience against emerging quantum computing threats.

Another significant advancement in quantum-resistant technology is Quantum Key Distribution (QKD). Rooted in quantum physics, QKD is a method for generating and sharing encryption keys that are impervious to decryption efforts, even by quantum computers. Quantum key distribution (QKD) protects data from being attacked by quantum computers and other powerful computational resources. It offers defence against existing attack techniques and, critically, against upcoming advancements in mathematics and quantum computing [25].

### 3.1 PQC Primitives:

Post Quantum Cryptography (PQC) primitives are the essential building blocks utilised to create cryptographic algorithms that are resistant to threats from quantum computers. These primitives are essential to many cryptographic operations, guaranteeing the confidentiality, integrity, and authenticity of the data. Key Exchange and Key Encapsulation Mechanisms, Digital Signatures, and Encryption are the 3 PQC primitives. Hardware post-quantum cryptographic primitives consist of four frequently used security components, i.e., Public Key Cryptosystem, Key Exchange, Oblivious transfer, and Zero-Knowledge Proof [26].

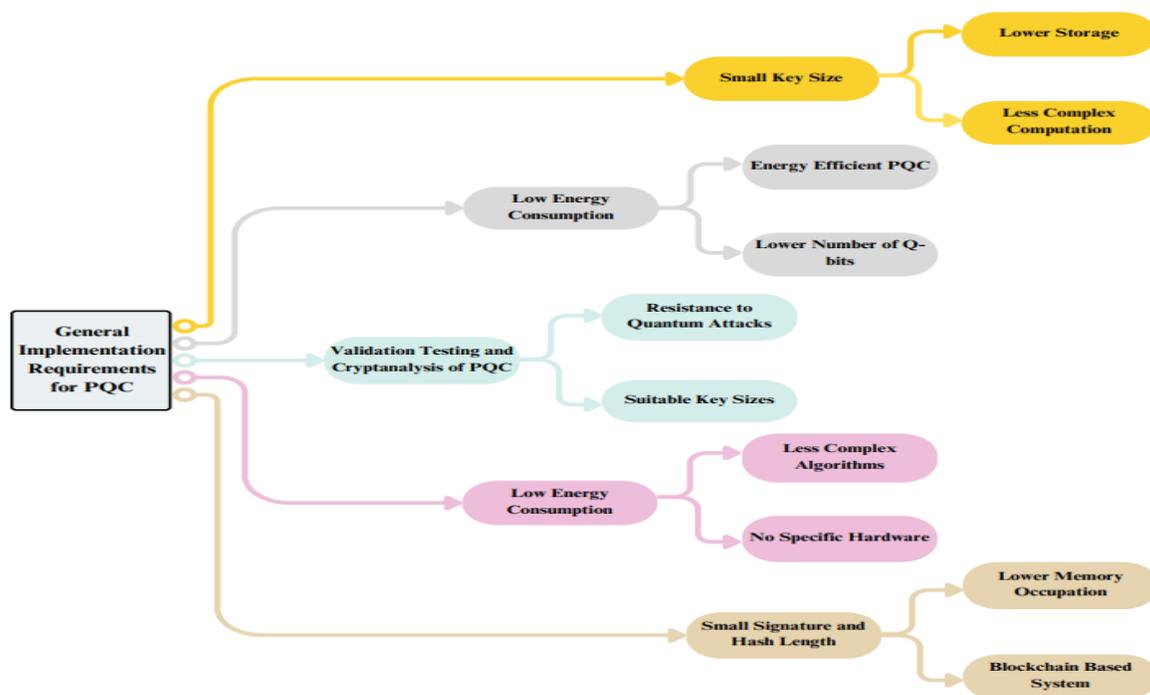

**Figure-3: PQC General Implementation Requirements [27]**

### 1.1 PQC General Implementation Requirements:

As quantum computing capabilities advance, the security of traditional cryptographic algorithms becomes increasingly precarious. Post-quantum cryptography offers a solution by developing algorithms resistant to quantum computer attacks. However, successful implementation requires adherence to general requirements, ensuring security, efficiency, and practicality. Figure 3 explores visual representation of the requirements for deploying effective post-quantum cryptographic systems in real-world settings.

### 1.2   IoT Perspective

In today's digital landscape, a vast number of interconnected IoT devices are seamlessly integrated across various sectors. The number of connected IoT devices is expected to be 29.42 billion by 2030. These interconnected IoT devices communicate with each other and use existing cryptographic algorithms like RSA, ECC, and AES for encryption. As quantum computers will break the existing cryptographic algorithms in the near future, massive amounts of data generated by interconnected IoT devices will be in danger. Fundamental building blocks of cryptography, key exchange, and encryption will be compromised. IoT networks are vulnerable to various attacks, including Sybil, eclipse, replay, side-channel, and false data injection. As a result, quantum-resistant post-quantum cryptography (PQC) needs to be integrated into IoT.

Major IoT applications such as smart homes, smart cities, healthcare, agriculture, smart grids, smart traffic light systems, and smart transportation systems usually make use of small and resource-constrained devices (belonging to Class 0, Class 1, or Class 2 devices with less than 10 KB of RAM or storage space and can have less than 100 KB of code in their flash memory) [28]. As a result, it is really hard to integrate PQC into resource-constrained IoT devices. ARM Cortex M3 or even M0 processors that are used in many IoT devices have a clock speed (execution performance) often as low as 8-24MHz and rising to 100-300MHz [29]. The PQM4 project suggests integrating performance metrics, such as code size, RAM consumption, and cycle counts, into NIST's PQC decision-making [29]. The assumption is that code and RAM utilisation won't rise much when moving from M4 to smaller devices (M3 or M0) [29]. A comparison of Falcon and Dilithium, which are NIST PQC digital signature schemes, reveals that while the Falcon code is 160KB and only needs 500B of RAM (with an 80KB code-size trade-off for 4–8K RAM), the Dilithium code is 12–20KB but uses 40–70KB of RAM [29]. As a result, Dilithium does not fit on these low-end resource-constrained IoT devices, even if it is 100 times better than Falcon. On the other hand, Kyber (NIST PQC KEMs) takes precedence due to its lower resource requirements.

## 2   Overview of HealthCare IoT

The integration of the Internet of Things (IoT) into healthcare systems has changed the landscape of patient care, data management, and medical infrastructure. IoT offers real-time monitoring, remote care, and data-driven insights by combining sensors, devices, and software, revolutionising healthcare delivery and patient outcomes. Various IoT sensors and devices serve different purposes, and each has a distinct role to play in contributing to diverse healthcare IoT applications.

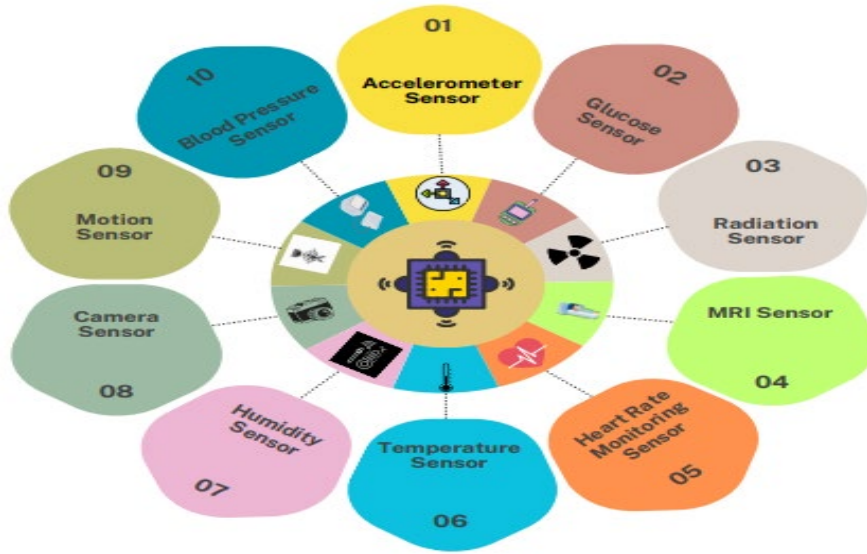

**Figure-4: Popular Healthcare IOT Sensors**

Figure 4 showcases the popular sensors commonly employed in healthcare IoT. These sensors play a vital role in collecting patient data, providing real time health updates in the healthcare IoT industry, and securely transmitting the data to the cloud through an IoT gateway.

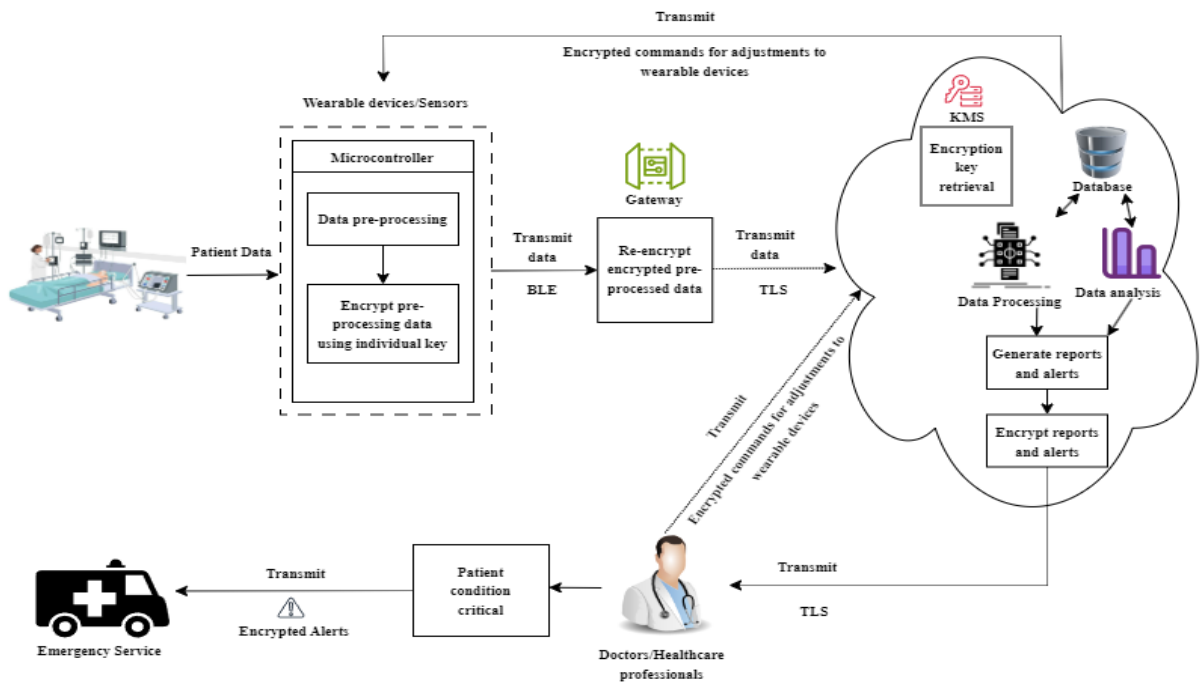

**Figure-5: IOT Based Smart Healthcare System**

These acquired data points are crucial inputs for doctors, allowing them to conduct thorough studies and provide detailed reports. Based on the report generated by the cloud, doctors or healthcare professionals can remotely adjust the IoT wearable device settings and send alerts to emergency services if the patient's condition is critical. The integration of an interconnected network of sensors, IoT devices, medical professionals, the IoT cloud, and emergency systems forms an IoT based smart healthcare system. In Figure 5 an IoT based smart healthcare system is given.

### 2.1 Cryptographic Applications in Healthcare IoT:

Cryptographic techniques are crucial in safeguarding the confidentiality and integrity of sensitive medical data in the quickly changing healthcare IoT (Internet of Things) sector. This section explores how cryptography ensures confidentiality, integrity, and availability (CIA) of medical data. Cryptographic solutions safeguard against unwanted access and preserve the reliability of healthcare services by encrypting data transmissions, confirming data integrity, and establishing secure communication channels.

Table-3: Cryptographic applications associated with the CIA triad for securing healthcare IoT environments.

| CIA | Application | Examples | Benefits |
| --- | --- | --- | --- |
| Confidentiality | Ensures encryption of medical data during transmission and storage to safeguard patient privacy [30]. | TLS, AES, ECC | Prevents unauthorised access, mitigates data breaches, ensures compliance with privacy regulations (e.g., HIPAA, GDPR). |
| Integrity | Utilises cryptographic hash algorithms to verify authenticity and integrity of transactions and medical information [30]. | SHA-256, HMAC | Detects unauthorised alterations, maintains data accuracy, enhances trust in medical records. |
| Availability | Ensures continuous availability and accessibility of medical records and equipment to authorised users [30]. | TLS, PKI | Facilitates prompt access to critical information, improves patient outcomes, enhances operational efficiency. |

### 2.2 Implications of Quantum Threat to Healthcare IoT:

To protect sensitive patient data transferred across networks, healthcare IoT devices frequently use traditional cryptographic encryption. Encryption algorithms, including RSA, AES, and ECC, are commonly used by healthcare IoT devices to provide secure data transmission and storage [31]. With the extraordinary computational power of quantum computers, traditional cryptographic algorithms that are widely being used in today's healthcare IoT devices will be vulnerable in the near future. In the future, with the quantum computer attack, hackers could hack healthcare IoT devices and access, read, and change private healthcare information, including medical records like patients' names, diagnoses, allergies, treatment histories, and others, which would compromise the confidentiality and integrity of healthcare data. Real-time health data like heart rate, blood sugar levels, and even brain activity will also become vulnerable due to the quantum computer attack. Hackers could take device control, such as pacemakers, insulin pumps, smart inhalers, and other critical IoT devices, which could lead to life-threatening consequences. There is also another threat known as "Harvest Now, Decrypt Later," in which cybercriminals seize encrypted copies of vital medical data that have a lengthy shelf life in order to launch an assault when quantum computing becomes accessible [32]. The implication of quantum threats goes beyond privacy in the healthcare IoT sector, potentially endangering human life. Figure 6 visually represents the quantum threats to IoT-based smart healthcare systems, identifying threats across different sections.

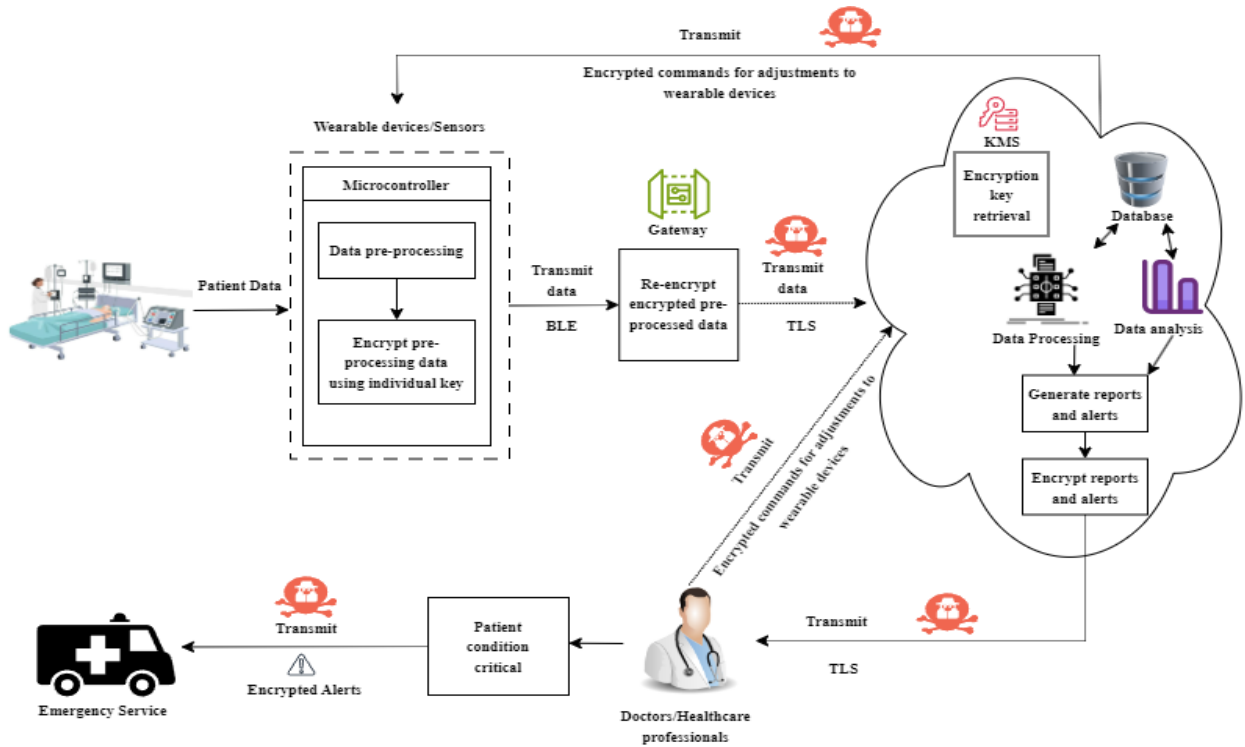

**Figure-6: Quantum Threat to IOT Based Smart Healthcare System**

Three use cases of IoT applications used in healthcare and impact of quantum threat on each of those use cases are given below:

### 2.2.1   Use Case-1: Remote Glucose Monitoring for Diabetics

Diabetic individuals may now regularly check their blood glucose levels without having to prick their fingers often thanks to remote glucose monitoring. This innovative approach is made possible through the use of Continuous Glucose Monitors (CGMs), compact wearable devices inserted beneath the skin that help to manage Type 1 or Type 2 diabetes with fewer fingerstick tests [33]. Every few minutes, these CGMs take a reading of the interstitial fluid's glucose data, which they then wirelessly send to a receiver or smartphone [33]. Unlike traditional blood glucose meters (BGMs), which offer a single measurement at a specific moment in time, CGMs provide real-time glucose readings, typically updated every five minutes. This translates to approximately 288 readings per day, offering users a comprehensive and continuous view of their glucose levels throughout the day and night [34]. While offering invaluable insights, CGMs operate with limited resources where RAM is 32-64 KB, ROM is 1-2 MB and clock speed is 16-32 MHz.

**Quantum Threats to Remote Glucose Monitoring for Diabetics IoT Devices**

Continuous glucose monitors (CGMs) provide remote glucose monitoring, which is a revolutionary development in diabetes care as it provides real-time information on blood glucose levels without requiring regular fingerstick testing. The security and integrity of patient data, however, are at danger due to the integration of this cutting-edge technology, which also creates vulnerabilities to quantum assaults. Weaknesses in the encryption algorithms used for the wireless transfer of glucose data from CGMs to receivers or smartphones might be exploited by quantum attackers. Adversaries might jeopardise patient privacy and security by intercepting this data and compromising the confidentiality of critical medical information, including real-time glucose levels.

Furthermore, strong quantum algorithms may be able to break the encryption techniques now used in CGMs. This flaw would make it possible for quantum attackers to crack encryption keys and have unauthorised access to patient data that CGMs store or communicate. Consequently, attackers may take part in illicit glucose reading tampering or manipulation, which might affect the precision of diabetes treatment and jeopardise patient care. In addition, CGMs are vulnerable to firmware flaws since they have limited resources, including RAM, ROM, and clock speed. Quantum attackers may take advantage of these flaws to compromise the usefulness and dependability of remote glucose monitoring for diabetics by gaining unauthorised access, inserting malware, or manipulating data within the CGMs.

With the advent of quantum computing, attackers possess the capability to intercept and decrypt CGM data, manipulate it, and subsequently transmit it back to the smartphone or cloud. Such manipulation of data could lead to healthcare providers administering incorrect treatments, thereby posing serious risks to patient safety and potentially endangering lives.

**Mitigating Quantum Threats Remote Glucose Monitoring for Diabetics IoT devices**

Implementing post-quantum cryptographic (PQC) algorithms and hardware-based security measures within Continuous Glucose Monitors (CGMs) significantly enhances their resilience against emerging quantum threats. PQC techniques, such as lattice-based or hash-based encryption, preserve patient privacy by ensuring that glucose data is securely sent wirelessly. Additionally, tamper-proof modules and quantum-resistant key storage techniques protect CGMs against unwanted access or data tampering [35]. Regular firmware upgrades, including security patches, are critical for fixing vulnerabilities and keeping CGMs resilient to possible quantum assaults. These procedures together assure the dependability and security of remote glucose monitoring systems, protecting patient data in the digital era.

### 2.2.2   Use Case-2: Smart Inhaler for Asthma Management in Healthcare IoT

Smart inhalers, when connected to smartphones, provide real-world data on patients' adherence patterns and inhaler technique during regular usage. While they can detect when a dosage is ready or activated, they cannot directly verify inhalation of the dose [36] . However, clinical research has demonstrated their ability to improve adherence and health outcomes, and they can send reminders for missed doses [36]. Several smart inhaler models are available or in development, including HeroTracker®, Doser™, INCA, RS01, and i-Neb [36]. Within smart inhalers, microcontrollers such as the QN9090 and K32W061 play critical roles in controlling device operations, analysing sensor data, and protecting patient information [37]. These microcontrollers also enable secure data transmission to the cloud, allowing healthcare professionals to remotely adjust inhaler settings based on patient requirements [37]. The K32W061, an ARM Cortex-M4 microcontroller, boasts a maximum computational power of 48MHz, along with 512 KB of RAM, 128 KB of ROM, and 640 KB of flash memory [38]. Similarly, the QN9090, also an ARM Cortex-M4 microcontroller, offers up to 48MHz computational power, 640 kB flash memory, up to 152 kB RAM, and 128 kB ROM [39]. These microcontrollers are integral to the functionality and security of smart inhalers, bridging the gap between patient care and technology.

**Quantum Threats to Smart Inhaler for Asthma Management IoT Devices**

The ARM Cortex-M4 microcontrollers found in smart inhalers for asthma management are vulnerable not only to quantum computing threats but also to side-channel attacks like Differential Power Analysis (DPA) and Fault Injection attacks [40]. DPA involves analysing power consumption patterns to extract cryptographic keys processed on the Cortex-M4, while Fault Injection attacks manipulate the device's hardware to induce errors and potentially extract information or alter its behavior [40]. These microcontrollers are vital for regulating device operations, analysing sensor data, and securely transmitting patient information to cloud-based platforms. During data transmission to cloud-based platforms, there's a risk of Man-in-the-Middle (MitM) attacks, where communication between the Cortex-M4 and other devices may be intercepted, allowing eavesdropping or data modification. Additionally, traditional cryptographic algorithms implemented in ARM Cortex-M4 microcontrollers are susceptible to quantum computing advancements. The emergence of quantum computers makes conventional encryption algorithms vulnerable, potentially leading to the decryption of encrypted patient data from smart inhalers. Consequently, patient data confidentiality and integrity are at risk.

These quantum threats extend beyond data security concerns, potentially impacting patient health, well-being, and trust in healthcare technologies. If the embedded microcontrollers are successfully attacked, patients may face incorrect treatment adjustments, compromised confidentiality, and psychological distress. Manipulated data could lead to misdiagnosis or delays in medical interventions, posing risks to patient safety. Addressing these threats requires robust security measures to safeguard patient data and ensure the integrity of asthma management processes.

**Mitigating Quantum Threats in Smart Inhaler Asthma Management IoT devices**

Mitigating quantum threats in smart inhalers for asthma management within IoT systems demands proactive measures to fortify patient data security and device integrity. This involves adopting post-quantum cryptography (PQC) algorithms such as Kyber and Dilithium, providing robust encryption capabilities resilient to quantum attacks [24]. Furthermore, enhancing ARM Cortex-M4 microcontroller security against side-channel attacks and implementing secure communication protocols like TLS, secure coding practices, and using well-tested libraries, regular security updates, and promptly patching vulnerabilities are imperative. Exploring quantum-resistant cryptography and ensuring continuous monitoring with timely security updates further fortify smart inhalers against evolving quantum threats, safeguarding patient health, well-being, and fostering trust in healthcare technologies.

### 2.2.3   Use Case-3: Heart Rate Monitoring

A heart rate monitoring system within the realm of healthcare IoT comprises a network of interconnected devices and sensors dedicated to continuously monitoring and tracking a patient's heart rate in real time. Standard components of such systems include wearables equipped with heart rate sensors alongside ancillary hardware like tablets, smartphones, or cloud-based platforms for data processing and analysis. Notably, Photoplethysmography (PPG) and Electrocardiogram (ECG or EKG) serve as the primary sensors utilised for heart rate monitoring [41, 42]. In terms of processing capabilities, heart rate monitoring systems commonly integrate microcontrollers from the STM32 and MSP430 series [43, 44]. For instance, STM32H5 microcontrollers feature Arm Cortex-M33 cores capable of operating at frequencies up to 250 MHz, accompanied by flash memory ranging from 128 Kbytes to 2 MB and up to 640 Kbytes of RAM [45]. On the other hand, MSP430 series microcontrollers typically offer 512 KB of flash memory and RAM capacities ranging from 125 bytes to 66 KB [46]. These microcontrollers play a pivotal role in managing data acquisition, processing, and transmission tasks within the heart rate monitoring system, ensuring real-time monitoring accuracy and efficiency.

**Quantum Threats to Heart Rate Monitoring IoT Devices**

The quantum threats confronting heart rate monitoring devices within healthcare Internet of Things (IoT) systems carry significant implications for patient well-being and overall healthcare outcomes. The vulnerability of conventional encryption protocols to decryption by advanced quantum algorithms, such as Shor's algorithm, jeopardises patient data security. This risk extends to raw photoplethysmography (PPG) and electrocardiogram (ECG) data, exposing sensitive details about heart function, sleep patterns, and emotional states. Breaches in confidentiality not only compromise patient privacy but also undermine trust in healthcare systems. Furthermore, the exploitation of vulnerabilities in widely used microcontrollers like the STM32 and MSP430 series by quantum adversaries amplifies the threat to patient health. Manipulated heart rate data, resulting from unauthorised access, could lead to erroneous diagnoses and ineffective treatment plans, potentially aggravating existing medical conditions or delaying necessary interventions. Inaccurate monitoring due to data manipulation may also result in missed opportunities for early detection of cardiac abnormalities or other health concerns, further compromising patient outcomes. Heart rate monitoring devices in healthcare IoT systems remain vulnerable to side-channel attacks, leveraging weaknesses in commonly used microcontrollers like STM32 and MSP430 [47]. These attacks exploit unintended information leakage through physical characteristics such as power consumption or electromagnetic radiation [48]. By analysing these patterns, attackers can infer sensitive data regarding patient health and device operations, potentially compromising the integrity of heart rate monitoring systems. The potential for large-scale Denial-of-Service (DoS) attacks adds another layer of risk, disrupting access to critical patient data precisely when it is needed most, such as during medical emergencies. Such disruptions could hinder healthcare professionals' ability to make timely and informed decisions, potentially compromising patient safety and exacerbating medical emergencies. Furthermore, the prospect of impersonation attacks enabled by quantum capabilities poses a significant threat to patient information integrity across the healthcare ecosystem. Unauthorised access to heart rate monitoring devices could lead to falsification of medical records or tampering with treatment protocols, undermining the trustworthiness of patient data and potentially endangering patient lives. In essence, the quantum threats to heart rate monitoring devices in healthcare IoT systems not only compromise data security but also directly impact patient health and safety.

**Mitigating Quantum Threats in Heart Rate Monitoring IoT Devices**

In light of the quantum threats faced by heart rate monitoring devices within healthcare IoT systems, it is crucial to implement tailored mitigation strategies. These strategies not only safeguard patient data but also ensure the integrity of healthcare outcomes. Specifically, for heart rate monitoring, proactive measures involve adopting Post-Quantum Cryptography (PQC) algorithms like Kyber, Dilithium, and NTRUEncrypt, as recommended by the National Institute of Standards and Technology (NIST) [49]. These robust algorithms provide long-term encryption resilience against quantum attacks, thereby maintaining the confidentiality of patient data transmitted by heart rate monitoring devices.

Furthermore, the utilisation of Quantum Key Distribution (QKD) technology establishes secure communication channels exclusively for heart rate monitoring devices [50]. This approach guarantees both the confidentiality and integrity of data transmission. To enhance security at the device level, hardware-based solutions such as tamper-proof modules and quantum-resistant key storage mechanisms offer additional layers of protection [35]. Additionally, continuous monitoring and threat detection mechanisms enable proactive identification and mitigation of potential quantum threats. By adhering to these practices, we uphold patient privacy and safeguard the integrity of healthcare outcomes in an increasingly quantum-enabled environment.

## 3   Mitigation Prospect and Challenges:

Fast-evolving quantum computing approaches continuously challenge traditional cryptography-enabled IoT applications in the healthcare industry. With its massive computational power, quantum computers will be able to break traditional cryptographic encryption algorithms used in healthcare IoT. However, with the advancement of quantum computers, the attackers are targeting the hardware part of the IoT devices that can affect the entire system. These attacks exploit physical vulnerabilities in the IoT hardware implementation rather than flaws in the mathematical structure of the cryptographic algorithms [51].

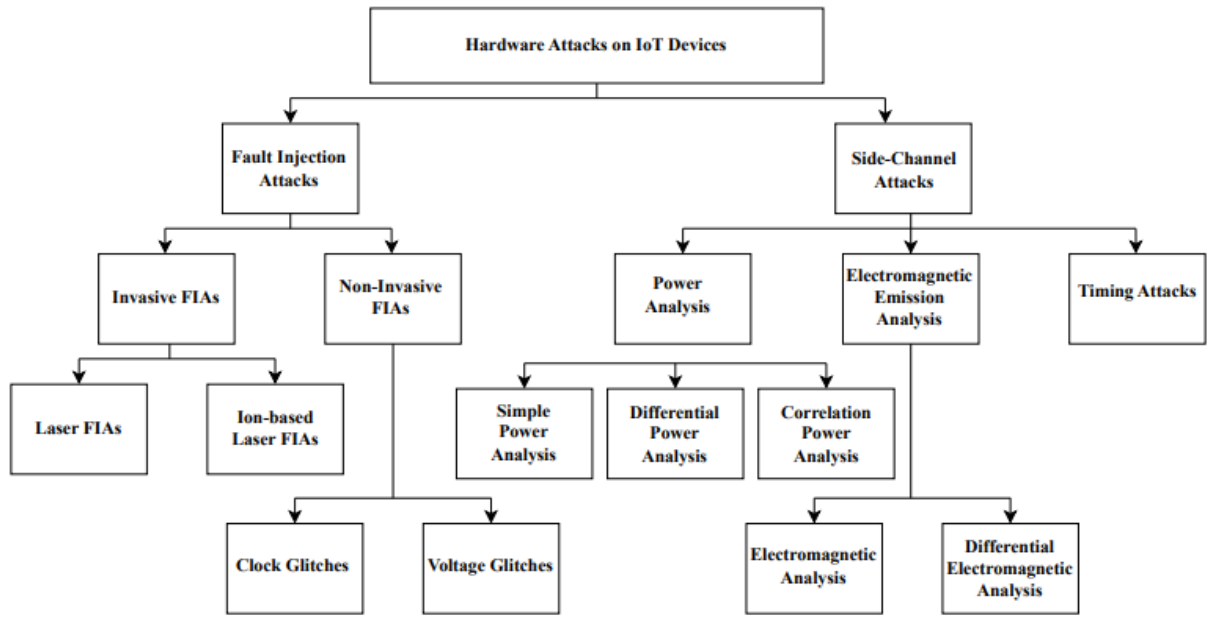

Figure-7: Hardware Attacks on IoT Devices [51]

Figure 7 presents a hierarchical representation illustrating hardware attacks on IoT devices enabled by advancements in quantum computing. To withstand the quantum computer attacks against healthcare IoT, Post Quantum Cryptographic schemes integration would be a valuable strategy. Code-based cryptography, Lattice-Based cryptography, Multivariate cryptography, and Hash-Based signatures are the four basic types of Post Quantum Cryptography [52]. Many organisations are trying to integrate these PQC algorithms on IoT networks to withstand quantum computer attacks in future. Furthermore, many IoT implementations, such as NB-IoT and LTE technologies, are currently undergoing standardisation processes to enhance their security. This process poses limitations on the integration of secure quantum alternatives into current IoT models [52, 60]. Post-quantum cryptography (PQC) schemes provide a feasible option for integration into current IoT/IT infrastructures, unlike quantum cryptography and quantum key distribution schemes, which demand specific and expensive equipment and primarily focus on key establishment; however, PQC schemes generally require greater memory and computational resources compared to traditional cryptographic solutions, presenting implementation hurdles for constrained IoT end nodes, such as low-performance microcontrollers with limited memory, even with traditional asymmetric cryptography such as RSA using 2K-bit keys [53]. Within the realm of post-quantum cryptography, lattice-based cryptosystems stand out as prime candidates for IoT devices, necessitating efficient polynomial-based multiplication calculations to minimize energy consumption and execution time, while similar optimisation challenges arise in other post-quantum cryptography mechanisms operating within finite fields specifically tailored for IoT devices [28].

### 3.1   Post Quantum Cryptographic Projects and Initiatives:

A number of research efforts and standardisation initiatives aiming at developing cryptographic solutions that can resist quantum challenges are being driven by post-quantum computing, which has emerged as a major field of study. Together with the development of standardised protocols, these efforts also include post-quantum cryptography projects.

**PQCrypto Project:** The PQCrypto project, funded by the European Union (EU), focused on investigating post-quantum cryptography for Internet communications, cloud computing, and low-power embedded devices. The pqm4 library, a result of the H2020 PQCRYPTO project, provides a practical solution customised for ARM Cortex-M4 microcontrollers, including various implementations of post-quantum key-encapsulation mechanisms and signature schemes, serving as a valuable benchmarking and testing framework specifically tailored for these microcontrollers [53].

**Open Quantum Safe (OQS) project:** The Open Quantum Safe (OQS) project, operating under the Linux Foundation's Post-Quantum Cryptography Alliance, is an open-source initiative dedicated to advancing quantum-resistant cryptography, providing quantum-resistant cryptographic algorithms through the liboqs C library and integrating these algorithms into various protocols and applications, including a provider for the widely used OpenSSL library, thus supporting both its own research efforts and the broader scientific community in addressing the challenges posed by quantum computing [54]. The OQS algorithm performance visualisations, derived from the OQS profiling project, offer crucial insights into the runtime, memory usage, and performance comparisons of Key Encapsulation Mechanism (KEM) and Signature (SIG) algorithms, vital for evaluating post-quantum cryptography's efficacy against quantum threats [55].

**PQClean project:** PQClean collaboratively develops refined implementations of post-quantum cryptographic schemes identified by the NIST post-quantum project, aiming for seamless integration with libraries like liboqs, support in higher-level protocols such as Open Quantum Safe, and inclusion in benchmarking frameworks like SUPERCOP, while serving as optimal starting points for architecture-specific optimisations, security evaluations, and formal verification processes [56].

**SAFECrypto project:** The SAFEcrypto project, funded by the EU and led by Queen's University of Belfast (United Kingdom), collaborated with partners from Switzerland, France, Germany, and Ireland to develop new post-quantum cryptosystems, focusing on lattice problems as a source of computational hardness [57].

NIST Initiatives: In December 2016, the National Institute of Standards and Technology (NIST) initiated a call for papers to standardise Post-Quantum Cryptography (PQC) algorithms. Among the 82 submission packages received, 69 met the requirements and were accepted. Following public review and feedback, seven finalists and eight alternative algorithms advanced to the third round in 2020. In 2022, NIST finalised the selection process, endorsing three digital signature algorithms, one Key Encapsulation Mechanism (KEM) algorithm for standardisation and four KEM algorithms proceeded to the fourth round [58]. After rigorous evaluation during the third round, NIST identified four candidate algorithms among them for most use cases, NIST recommended implementing two primary algorithms: CRYSTALS-KYBER for key establishment and CRYSTALS-Dilithium for digital signatures [58].

IETF initiatives:  The Internet Engineering Task Force (IETF), in collaboration with the Crypto Forum Research Group (CFRG), is working on Internet-Drafts related to quantum cryptography. These drafts cover topics such as post-quantum cryptosystems for TLS and extending protocols like IKEv2 to be post-quantum resistant [59]. IETF is actively involved in updating security protocols for post-quantum cryptography (PQC) to address quantum threats. Key initiatives include the formation of working groups like PQUIP, focused on integrating PQC algorithms into existing protocols [61]. These efforts aim to ensure interoperability, security, and resilience in internet communication protocols such as TLS, SSH, and HTTPS. Through collaboration across academia and industry, the IETF seeks to architect a smooth transition to PQC, safeguarding internet infrastructure against emerging quantum vulnerabilities.

### 3.2   PQC Implementation Challenges in Healthcare IoT:

The rapid progress of quantum computing poses a significant risk to the security of traditional cryptographic methods used in today's healthcare Internet of Things (IoT) devices. While post-quantum cryptography (PQC) presents a solution to this impending threat, its implementation in healthcare IoT environments introduces distinctive challenges. Adapting post-quantum cryptographic solutions to the specific needs of the healthcare sector and mitigating the vulnerabilities posed by quantum computing demand careful consideration and specialized approaches. Therefore, implementing post-quantum cryptography in healthcare IoT settings requires addressing these unique challenges effectively. Figure 8 illustrates the key challenges in implementing post-quantum cryptography (PQC) within Healthcare IoT environments.

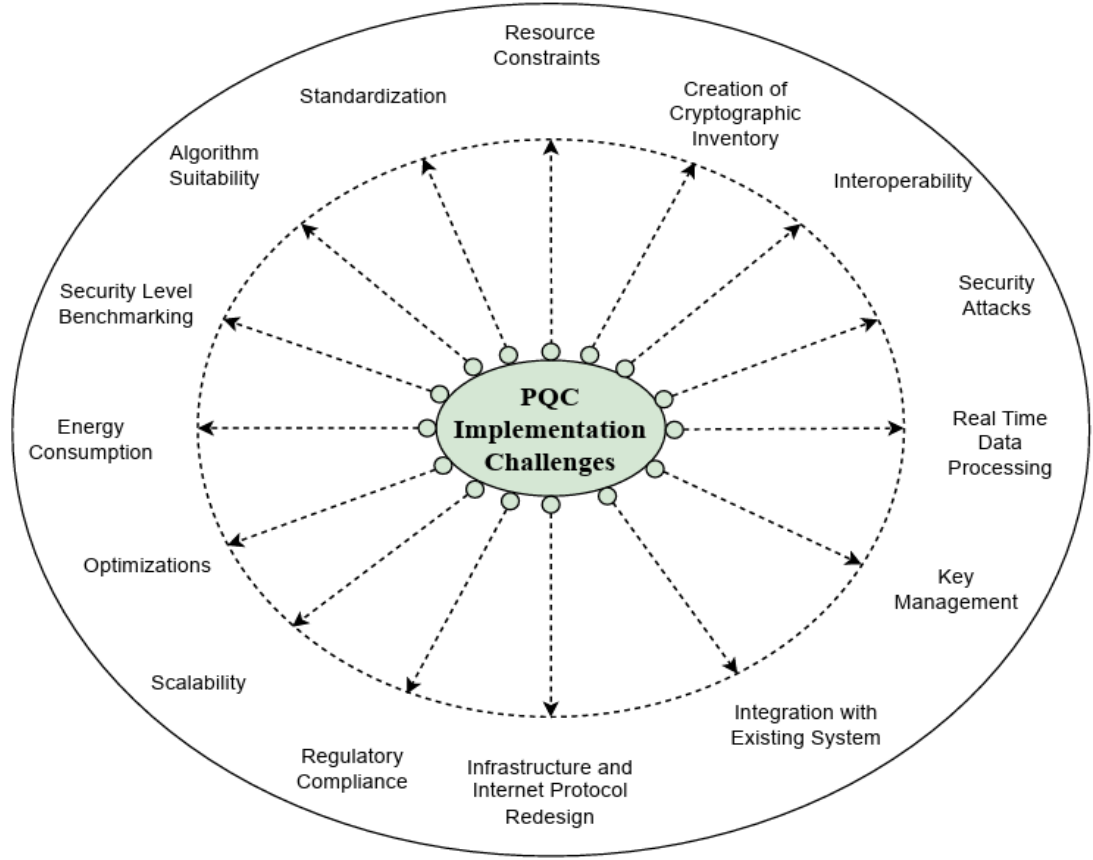

Figure-8: PQC Implementation Challenges in Healthcare IoT

## 4 Conclusion:

The quantum threat poses a significant risk to the healthcare Internet of Things (IoT), jeopardising the security of patient data and devices. We investigated post-quantum cryptography (PQC) in this chapter as a possible means of protection, evaluating its appropriateness for medical devices with limited resources. Despite the presence of obstacles in implementation, the ongoing mitigation projects show potential. Collaboration among healthcare providers, manufacturers, and regulators is essential for ensuring the security of healthcare IoT. Ongoing PQC research and global collaboration are crucial. To ensure the protection of patient privacy, device functionality, and trust in the future of digital healthcare, it is crucial to take proactive measures at present.

Securing Healthcare IoT   23